\documentclass[12pt]{iopart}
\eqnobysec

\usepackage{iopams,graphicx,amssymb}

\begin{document}

\title[Anomalous scaling of a passive vector field in $d$ dimensions]
{Anomalous scaling of a passive vector field in $d$ dimensions:
Higher-order structure functions}

\author{L~Ts~Adzhemyan, N~V~Antonov, P~B~Gol'din and M~V~Kompaniets}

\address{Department of Theoretical Physics, St.~Petersburg University,
Uljanovskaja 1, St.~Petersburg, Petrodvorez, 198504 Russia}

\ead{nikolai.antonov@pobox.spbu.ru}

\begin{abstract}
The problem of anomalous scaling in the model of a transverse vector field
$\theta_{i}(t,{\bf x})$ passively advected by the non-Gaussian, correlated in
time turbulent velocity field governed by the Navier--Stokes equation, is
studied by means of the field-theoretic renormalization group and operator
product expansion. The anomalous exponents of the $2n$-th order structure
function
$S_{2n}(r) = \langle [\theta(t,{\bf x}) - \theta (t,{\bf x}+{\bf r})]^{2n}
\rangle $, where $\theta$ is the component of the vector field parallel to
the separation ${\bf r}$, are determined by the critical dimensions of the
family of composite fields (operators) of the form
$(\partial\theta\partial\theta)^{2n}$, which mix heavily in renormalization.
The daunting task of the calculation of the matrices of their critical
dimensions (whose eigenvalues determine the anomalous exponents) simplifies
drastically in the limit of high spatial dimension, $d\to\infty$.
This allowed us to find the leading and correction anomalous exponents
for the structure functions up to the order $S_{56}$.
They reveal intriguing regularities, which suggest for the anomalous
exponents simple ``empiric'' formulae that become practically exact for
$n$ large enough. Along with the explicit results for smaller $n$, they
provide the full description of the anomalous scaling in the model.
\end{abstract}


\maketitle

\section{Introduction} \label{sec:Intro}

In the past two decades, much attention has been attracted by turbulent
advection of passive scalar fields; see the review paper \cite{FGV} and
references therein. Being of practical importance in itself, the problem
of passive advection can be viewed as a starting point for studying
intermittency and anomalous scaling in the fluid turbulence on the whole
\cite{Legacy}. Most progress was achieved for the so-called Kraichnan's
rapid-change model, in which the advecting velocity field $v_{i}(x)$
with $x = \{t,{\bf x}\}$ is modelled by a Gaussian statistics with vanishing
correlation time and prescribed correlation function
$\langle vv\rangle \propto \delta(t-t') k^{-d-\xi}$, where $k$ is the wave
number, $d$ is the dimension of space and $\xi$ is an arbitrary exponent with
the most realistic (Kolmogorov) value $\xi=4/3$. The structure functions
of the advected scalar field $\theta (x)$ in the inertial range
demonstrate anomalous scaling behaviour:
\begin{eqnarray}
S_{2n}(r) = \big\langle
[ \theta (t,{\bf x}) - \theta (t,{\bf x'})]^{2n} \big\rangle \propto
r^{n(2-\xi)}\, (r/L)^{\Delta_{n}},
\label{Ano}
\end{eqnarray}
that is, singular dependence on the separation $r=|{\bf r}|$ (where
${\bf r} = {\bf x}-{\bf x'}$) and
on the integral turbulence scale $L$, characterized by an infinite set of
the exponents $\Delta_{n}$. Within the framework of the so-called zero-mode
approach, these exponents were calculated in the leading order of the
expansions in $\xi$ \cite{GK} and $1/d$ \cite{Falk1}:
\begin{equation}
\Delta_{n}= -2n(n-1)\xi/(d+2)+O(\xi^{2}) = -2n(n-1)\xi/d+O(1/d^{2}).
\label{HZ3}
\end{equation}

In \cite{RG} and subsequent papers, the field theoretic renormalization
group (RG) and the operator product expansion (OPE) were applied to
Kraichnan's model; see \cite{JphysA} for a review and the references.
In that approach, the anomalous scaling emerges as a consequence of the
existence in the corresponding OPE of certain composite fields (``operators''
in the quantum-field terminology) with {\it negative} dimensions, which are
identified with the anomalous exponents $\Delta_{n}$. This allows one to
construct a systematic perturbation expansion for the anomalous exponents
and to calculate them up to the orders $\xi^{2}$ \cite{RG} and $\xi^{3}$
\cite{cube}. Besides the calculational efficiency, an important advantage
of the RG+OPE approach is its relative universality: it can also be applied
to the case of finite correlation time or non-Gaussian advecting fields.
For passively advected {\it vector} fields, any calculation of the exponents
for higher-order correlations calls for the RG techniques already in the
$O(\xi)$ approximation.

In this paper, we study anomalous scaling of a passive vector quantity,
advected by a non-Gaussian velocity field, governed by the stirred
Navier--Stokes (NS) equation. For the rapid-change velocity ensemble,
a similar model was introduced and thoroughly studied in
\cite{LA}--\cite{Novikov}; effects of finite correlation time and weak
anisotropy were studied in \cite{Sh}. Before explaining our motivations,
which follow the same lines as those of refs. \cite{LA,vector,matrix},
let us discuss the definition of the model.

We confine ourselves with the case of transverse (divergence-free) passive
$\theta_{i} (x)$ and advecting $v_{i}(x)$ vector fields.
Then the general advection-diffusion equation has the form
\begin{eqnarray}
\nabla _t\theta_{i} - {\cal A} (\theta_{k}\partial_{k}) v_{i} +
\partial_{i} {\cal P} = \kappa_0\partial^{2} \theta_{i} + \eta_{i},
\qquad
\nabla_t \equiv \partial _t + (v_{k}\partial_{k}) ,
\label{1}
\end{eqnarray}
where $\nabla_t$ is the Lagrangian (Galilean covariant) derivative,
${\cal P}(x)$ is the pressure, $\kappa_0$ is the diffusivity,
$\partial^{2}$ is the Laplace operator and $\eta_{i}(x)$ is a transverse
Gaussian stirring force with zero mean and covariance
\begin{equation}
\langle \eta_{i}(x) \eta_{k}(x')\rangle = \delta(t-t')\, C_{ik}(r/L).
\label{2}
\end{equation}
The parameter $L$ is an integral scale related to the stirring, and
$C_{ik}$ is a dimensionless function, finite at $r=0$ and rapidly decaying
for $r\to\infty$; its precise form is unimportant. Due to the transversality
conditions $\partial_{i}\theta_{i} = \partial_{i}v_{i}=0$, the pressure
can be expressed as the solution of the Poisson equation,
\begin{equation}
\partial^{2} {\cal P} = ({\cal A} - 1) \,
\partial_{i} v_{k}  \partial_{k} \theta_{i}.
\label{Pois}
\end{equation}
Thus the pressure term makes the dynamics (\ref{1}) consistent with the
transversality. The amplitude factor ${\cal A}$ in front of the
``stretching term'' $(\theta_{k}\partial_{k}) v_{i}$ is not fixed by the
Galilean symmetry and thus can be arbitrary. Such general ``${\cal A}$ model''
was introduced and studied in refs. \cite{amodel}. The most popular special
case ${\cal A}=1$, where the pressure term disappears, corresponds to
magnetohydrodynamic turbulence. It was studied earlier in numerous papers;
see e.g. refs. \cite{MHD} and references therein.

In earlier studies, the velocity field in (\ref{1}) was described by the
Kraichnan's rapid-change model. In this paper, we employ the stochastic
NS equation:
\begin{equation}
\nabla _t v_i=\nu _0\partial^{2} v _i-\partial _i {\wp}+f_i,
\label{1.1}
\end{equation}
where $\nabla _t$ is the same Lagrangian derivative, ${\wp}$ and $f_i$
are the pressure and the transverse random force per unit mass. We assume
for $f$ a Gaussian distribution with zero mean and correlation function
\begin{equation}
\big\langle f_i(x)f_j(x')\big\rangle = \frac{\delta (t-t')}{(2\pi)^{d}}\,
\int_{k\ge m} d{\bf k}\, P_{ij}({\bf k})\, d_f(k)\, \exp \big[{\rm i}{\bf k}
\left({\bf x}-{\bf x}'\right)\big] ,
\label{1.2}
\end{equation}
where $P_{ij}({\bf k}) =\delta _{ij}  - k_i k_j / k^2$ is the transverse
projector, $d_f(k)$ is some function of $k\equiv |{\bf k}|$ and model
parameters. The momentum $m=1/L$, the reciprocal of the integral scale
$L$ related to the velocity, provides IR regularization. For simplicity,
we do not distinguish it from the integral scale related to the scalar
noise in (\ref{2}).

The standard RG formalism is applicable to the problem (\ref{1.1}),
(\ref{1.2}) if the correlation function of the random force is chosen
in the power form \cite{DeDom}
\begin{equation}
d_f(k)=D_0\,k^{4-d-y},
\label{1.9}
\end{equation}
where $D_{0}>0$ is the positive amplitude factor and the exponent
$0<y\le 4$ plays the role of the RG expansion parameter. The
most realistic value of the exponent is $y=4$: with an
appropriate choice of the amplitude, the function (\ref{1.9}) for
$y\to4$ turns to the delta function, $d_f(k) \propto
\delta({\bf k})$, which corresponds to the injection of energy to
the system owing to interaction with the largest turbulent eddies;
for a more detailed justification see e.g. \cite{turbo,Book3}.

In this paper we consider the model (\ref{1}) without the stretching term,
that is, ${\cal A}=0$. Being formally a special case of the general
${\cal A}$ model, it appears exceptional in a few respects and requires
special attention \cite{LA}--\cite{Sh}.

The feature specific only of the ${\cal A}=0$ is the symmetry with respect
to the shift $\theta\to\theta+{\rm const}$, because only {\it derivatives}
of the field $\theta$ enter the equation (\ref{1}). The quantities of
interest are the structure functions (\ref{Ano}), in which $\theta$ should be
understood as the component of the vector field parallel to the separation,
$\theta= \theta_{i} r_{i}/r$: in contrast to ordinary correlation functions,
they are also invariant with respect to the shift. As a consequence, all
the composite operators that enter the corresponding OPE, should also be
invariant, that is, built only of the derivatives of $\theta$.

For the scalar problem, the operator that determines the leading term of
the inertial-range asymptotic behaviour (\ref{Ano}), is unique: it has
the form $(\partial _{i}\theta  \partial_{i} \theta)^{n}$, the $n$-th power
of the local dissipation rate of the scalar field fluctuations, and its
critical dimension gives the anomalous exponent $\Delta_{n}$ is (\ref{HZ3});
see \cite{RG}--\cite{cube} for the detailed discussion.

For the vector case one can construct many scalar operators of the form
$(\partial\theta)^{2n}$ for a given $n$, and in order to find the
corresponding {\it set} of exponents and to identify the leading
contribution, one has to consider the renormalization of the whole family,
which implies the {\it mixing} of individual operators \cite{LA,vector}.
Renormalization of families of mixing composite fields and calculation of the
corresponding matrices of critical dimensions (whose eigenvalues give the
desired anomalous exponents) its rather cumbersome and labor-consuming task,
which should be solved separately for different families (in the case at
hand, for different $n$). Thus, at first sight, there is no hope to derive
simple explicit expressions for the anomalous exponents, similar to
(\ref{HZ3}) in the scalar case.

In this respect, the ${\cal A}=0$ case of the model (\ref{1}) resembles the
nonlinear stirred NS equation, where the inertial-range behavior of structure
functions is believed to be related with the Galilean-invariant (and hence
built of the velocity gradients) operators, which mix heavily in
renormalization; see \cite{turbo} and references therein. In that case, the
full solution is not yet obtained even for the relatively simple case of the
family that includes the square of the energy dissipation rate \cite{TMF}.

Thus the ${\cal A}=0$ vector model (\ref{1}) is of special interest:
it also involves the problem of mixing, but now the problem is not a
hopeless one: the leading terms are determined by finite families of
composite operators, namely, those of the form $(\partial\theta)^{2n}$
with all possible contractions of vector indices, and such families with
a given $n$ are closed with respect to the renormalization \cite{LA,vector}.
What is more, for low values of $d$ there are linear relations between the
operators, which reduce drastically the number of independent monomials
\cite{Novikov}. As a result, the leading anomalous exponents for $d=2$
were calculated to the order $O(\xi^{2})$ for all $n$, and for $d=3$ --
to the order $O(\xi)$ for $n\le 9$ \cite{Novikov}. Crucial simplifications
also take place in the limit $d\to\infty$ \cite{vector}: like for the scalar
Kraichnan's model, the anomalous exponents decay as $O(1/d)$ for large $d$,
so that the anomalous scaling disappears at $d=\infty$; cf. (\ref{HZ3}).
In order to find the leading exponents (and the closest corrections) it
is sufficient to consider some special subset of the whole family
$(\partial\theta)^{2n}$, and the corresponding matrix of critical dimensions
can be built by a simple algorithm \cite{vector}. This allowed us to find
all the negative dimensions in the leading order of the double expansion in
$\xi$ and $1/d$ for $n$ as large as $n=28$, which gives the anomalous
exponents and the close corrections for the structure functions up to
$S_{56}$ \cite{matrix}. All those results, however, refer to the Gaussian
velocity field.

For very large $n$ the calculations become too labor- and time consuming
(mostly due to the diagonalization of the matrices), but they are not
necessary: the large-$n$ results suggest some simple empiric explicit
expressions for the leading, next-to-leading, {\it etc}, anomalous exponents,
which become practically exact as $n$ increases. Along with the explicit
answers for smaller $n$, this gives the complete description of the anomalous
scaling in the vector model for all $n$ and large $d$ \cite{matrix}.

It should be emphasized that the study of the large-$d$ behaviour of the
fluid turbulence is by no means of only academic interest. It is related to
the old idea of the expansion in $1/d$, which has repeatedly been introduced
in various contexts \cite{OK}--\cite{infty}.

The problem is that the ordinary perturbation theory for the stirred
(stochastic) NS equation (that is, the perturbation
expansion in the nonlinearity) is in fact an expansion in the Reynolds
number, a parameter which tends to infinity for the fully developed
turbulence. A similar problem is well known in the theory of critical state,
where it is solved by means of the RG techniques; see e.g. \cite{Book3}.
The RG allows one to rearrange (to sum up) the plain perturbation series
and to replace them with the famous $\varepsilon$ expansion, where
$\varepsilon=4-d$ is the deviation of the spatial dimension $d$ from its
upper critical value $d=4$. The turbulence (or, better to say, the
corresponding stochastic models), has no upper critical dimension, and the
RG expansion parameter has completely different meaning. As already mentioned,
in Kraichnan's model its role is played by the exponent $\xi$, while in the
RG approach to the stirred NS equation its analog is the exponent $y$ in
the correlator of the stirring force; see section~\ref{sec:QFT}.
The results of the RG analysis of this model are reliable and internally
consistent for asymptotically small $\xi$ or $y$, while the possibility
of their extrapolation to the physical finite values, and thus their
relevance for the real fluid turbulence, is sometimes called in question;
see the discussion and the references in \cite{infty,China}.

One can hope that in the limit $d\to\infty$ intermittency and anomalous
scaling disappear or acquire a simple ``calculable'' form and the
finite-dimensional turbulence can be studied within the expansion around
this ``solvable'' limit \cite{FFR}. Indeed, it was argued (on the basis of
a certain SDE-motivated ansatz for dissipative terms) that the Kolmogorov
theory becomes exact and the multiscaling indeed disappears for $d=\infty$
\cite{YakhotD}, as it also happens for the Obukhov--Kraichnan model \cite{OK}.
What is more, for the latter it was possible to find the $O(1/d)$
contribution to the anomalous exponents \cite{Falk1}; see the last
expression in (\ref{HZ3}). However, the systematic expansion in $1/d$ has not
been yet constructed for that model, let alone the stochastic NS equation.

It was suggested in \cite{Anton,infty} that new progress can be achieved
by combining the large-$d$ limit with the RG approach and the expansion
in $\varepsilon$. In particular, it was noticed \cite{vector,Anton} that
taking the limit $d\to\infty$ leads to serious simplifications in the RG
calculations, especially for composite operators. In a very important paper
\cite{Anton}, scaling dimensions of all the powers of the local energy
dissipation rate for the NS problem were calculated for $d=\infty$ to
first order in $y$, the problem that looks unfeasible for finite $d$;
see the discussion in \cite{TMF}.

Thus the drastic simplifications that occur in our vector model with the
turbulent mixing provided by the NS velocity field and the simple explicit
results for the anomalous dimensions in the leading order of the double
expansion in $y$ and $1/d$, give a new strong support to the idea of the
large-$d$ expansion.

The plan of the paper is as follows. In sec.~\ref{sec:QFT} we discuss
the field theoretic formulation of our stochastic problem and its
renormalization. We show that the corresponding RG equations have an IR
attractive fixed point, which implies existence of IR scaling behaviour
for various correlation functions. In sec.~\ref{sec:OPE}, inertial-range
anomalous scaling of the structure functions is studied by means of the
OPE and the role of the operators $(\partial\theta)^{2n}$ is clarified.
In sec.~\ref{sec:d} we discuss the renormalization of those operators and
the simplifications that occur in the limit of large $d$. Examples are
given of the matrices of critical dimensions for a few families of those
operators. In sec.~\ref{sec:n} the leading and correction anomalous
exponents, determined by the eigenvalues of those matrices, are presented
and the regularities that they reveal are discussed. These regularities
suggest for the eigenvalues some simple ``empiric'' formulae that become
practically exact for $n$ large enough. Along with the explicit results
obtained for smaller $n$, they provide the full description of the anomalous
scaling in the present model. Sec.~\ref{sec:Conc} is reserved for a brief
conclusion.

\section{Field theoretic formulation, renormalization and RG equations}
\label{sec:QFT}

According to the general theorem (see e.g. \cite{Book3}), the full-scale
stochastic problem (\ref{1})--(\ref{1.9}) for ${\cal A}=0$ is equivalent
to the field theoretic model of the doubled set of fields
$\Phi=\{v,v',\theta,\theta'\}$ with the action functional
\begin{equation}
{\cal S} (\Phi )= {\cal S}_{v}({\bf v}', {\bf v}) +
\theta' D_{\theta} \theta'/2 +
\theta' \left\{ -\nabla_{t}+ \kappa_{0} \partial^{2} \right\} \theta,
\label{action}
\end{equation}
where $D_{\theta}$ is the correlation function (\ref{2}) of the random
noise $f$ in (\ref{1}) and $S_{v}$ is the action for the problem
(\ref{1.1})--(\ref{1.9}):
\begin{equation}
{\cal S}_{v}({\bf v}', {\bf v}) =
v 'D_{v}v'/2+ v'\left\{-\nabla_t + \nu_0 \partial^{2} \right\}v ,
\label{actionV}
\end{equation}
where $D_{v}$ is the correlation function (\ref{1.2}) of the random force
$f_{i}$. All the integrations over $x=\{t,{\bf x}\}$ and summations
over the vector indices are understood. The auxiliary vector fields
$v',\theta'$ are also transverse,
$\partial_{i}v_{i}'=\partial_{i}\theta_{i}'=0$, which allows to omit the
pressure terms on the right-hand sides of expressions (\ref{action}),
(\ref{actionV}), as becomes evident after the integration by parts.
For example,
\[ \int dt \int d{\bf x} \ v_{i}'\partial_{i} {\wp} = - \int dt
\int d{\bf x} \ {\wp} (\partial_{i}v_{i}') =0 . \]
Of course, this does not mean that the pressure contributions are
unimportant: the fields $v',\theta'$ act as transverse projectors and
select the transverse parts of the expressions to which they are contracted.

The role of the coupling constants is played by the two parameters
$g_{0}\equiv D_{0}/\nu_0^3$ and $u_{0} = \kappa_{0}/\nu_0$, the analog of
the inverse Prandtl number in the scalar case. By dimension,
\begin{equation}
g_{0} \propto \Lambda^{y} \quad {\rm and} \quad u_{0} \propto \Lambda^{0},
\label{Lambda}
\end{equation}
where $\Lambda$
is the characteristic ultraviolet (UV) momentum scale. Thus the model
(\ref{action}), (\ref{actionV}) becomes logarithmic (both the coupling
constants become dimensionless) at $y=0$, and the UV divergences manifest
themselves as poles in $y$.

The renormalization and RG analysis of the model (\ref{action}),
(\ref{actionV}) are similar to that of the scalar advection by the NS
velocity field \cite{AVH,NSpass}, and here we discuss them only briefly.
Dimensional analysis, augmented with symmetry considerations (Galilean
invariance and the symmetry with respect to the shift
$\theta\to\theta+{\rm const}$) shows that superficial UV divergences are
present only in the 1-irreducible Green functions $\langle v'v \rangle$
and $\langle\theta'\theta\rangle$, and the corresponding counterterms have
the forms $v'\partial^{2}v$ and $\theta'\partial^{2}\theta$. They can be
reproduced by multiplicative renormalization of the parameters
\begin{equation}
\nu_0 = \nu Z_{\nu}, \quad \kappa_0 = \kappa Z_{\kappa}, \quad
g_{0} = g \mu^{y} Z_{g}, \quad Z_{g} =  Z_{\nu}^{-3};
\label{mult}
\end{equation}
no renormalization of the fields $\Phi$ and the IR scale $m$ is needed.
Here $\nu$, $g$, $\kappa$ are
renormalized analogs of the bare parameters $\nu_{0}$, $g_{0}$, $\kappa_{0}$
and the reference scale $\mu$ is an additional parameter of the renormalized
theory. The last relation in (\ref{mult}) follows from the absence of
renormalization of the amplitude $D_{0}=g_{0}\nu_{0}^{3}= g\mu^y \nu^3$ in
the first term of the action (\ref{actionV}). The renormalization constants
$Z_{i}=Z_{i}(g,u,d,y)$ absorb all the UV divergences, so that the Green
functions are UV finite (that is, finite at $y=0$) when expressed
in renormalized parameters.

The one-loop calculation gives:
\begin{eqnarray}
Z_{\nu}=1 - g \bar S_{d}\,  \frac{(d-1)}{4(d+2)} \, \frac{1}{y} +O(g^{2}),
\nonumber \\
Z_{\kappa}= 1 - g \bar S_{d} \, \frac{(d^{2}-3)}{2d(d+2)} \,
\frac{1}{yu(u+1)} +O(g^{2}),
\label{Z}
\end{eqnarray}
where  $\bar S_{d}=S_{d}/(2\pi)^{d}$ and $S_d\equiv 2\pi^{d/2}/\Gamma(d/2)$
is the surface area of the unit sphere in $d$-dimensional space.
Of course, due to the passivity of the field $\theta$, the constant
$Z_{\nu}$ is the same as in the model (\ref{actionV}).

Since the model is multiplicatively renormalizable, the RG equations can be
derived in the standard manner; see e.g. \cite{Book3}. The RG equation for
a certain renormalized Green function $G^{R}=\langle \Phi\dots\Phi\rangle$
has the form
\begin{eqnarray}
\left\{ {\cal D}_{\mu} - \gamma_{\nu} {\cal D}_{\nu} +
\beta_{g}\partial_{g}+ \beta_{u}\partial_{u} \right\} G^{R}=0.
\label{RG}
\end{eqnarray}
Here ${\cal D}_{s} = s\partial_{s}$ for any variable $s$, $u=\kappa/\nu$ and
the RG functions (the $\beta$ functions and the anomalous dimensions
$\gamma$) are defined as
\begin{equation}
\gamma_{F} = \widetilde{\cal D}_{\mu} \ln Z_{F}
\label{gamma}
\end{equation}
for any quantity $F$ and
\begin{eqnarray}
\beta_{g} = \widetilde{\cal D}_{\mu} g = g[-y+3\gamma_{\nu}],
\quad
\beta_{u} = \widetilde{\cal D}_{\mu} u = u[\gamma_{\nu}-\gamma_{\kappa}],
\label{beta}
\end{eqnarray}
where $\widetilde{\cal D}_{\mu}$ is the operation ${\cal D}_{\mu}$ at fixed
bare parameters and the second relations in (\ref{beta}) follow from the
definitions and the relations (\ref{mult}). It remains to note that the
differential operator in (\ref{RG}) is nothing but $\widetilde{\cal D}_{\mu}$
expressed in renormalized variables.

From (\ref{Z}) one obtains the following explicit one-loop expressions for
the anomalous dimensions:
\begin{eqnarray}
\gamma_{\nu}= g \bar S_{d}\,  \frac{(d-1)}{4(d+2)} +O(g^{2}),
\nonumber \\
\gamma_{\kappa}= g \bar S_{d} \, \frac{(d^{2}-3)}{2d(d+2)} \,
\frac{1}{u(u+1)} +O(g^{2}).
\label{explicit}
\end{eqnarray}

It is well known that IR asymptotic behaviour of a multiplicatively
renormalizable field theory is governed by IR attractive fixed points of the
corresponding RG equations. Their coordinates are found from the requirement
that all the $\beta$ functions vanish; in our case, $\beta_{g}=\beta_{u}=0$.
From the explicit expressions (\ref{beta}), (\ref{explicit}) for $\beta_{g}$
it follows that the model (\ref{actionV}) has a nontrivial fixed point
\begin{eqnarray}
g_{*} \bar S_{d} = y\, \frac{4(d+2)}{3(d-1)} + O(y^{2})
\label{fix1}
\end{eqnarray}
which is positive and IR attractive ($\partial_{g}\beta_{g}>0$) for $y>0$
(of course, this fact is well known, see e.g. \cite{turbo,Book3}). Then
from the equation $\beta_{u}=0$ and the explicit expressions (\ref{beta}),
(\ref{explicit}) and (\ref{fix1}) one obtains
\begin{eqnarray}
u_{*}(u_{*}+1)  =  \frac{2(d^{2}-3)}{d(d-1)} + O(y).
\label{fix2}
\end{eqnarray}
We are interested in the positive root of the equation (\ref{fix2}) which
exists for $d^{2}>3$. It is easily checked that this fixed point is IR
attractive ($\partial_{u}\beta_{g}=0$, $\partial_{u}\beta_{u}>0$).

Existence of an IR attractive fixed point in the physical range of parameters
($u_{*}>0$, $g_{*}>0$) means that the Green functions of the full model
(\ref{action}), (\ref{actionV}) exhibit self-similar (scaling) behaviour in
the IR asymptotic range. The corresponding critical dimensions
$\Delta[F]=\Delta_{F}$ of the basic fields and parameters $F$ are found in
the standard way (see e.g. \cite{turbo,Book3}) and especially \cite{NSpass}
for the analogous scalar problem) and have the forms
\begin{eqnarray}
\Delta[v^{n}]=n\Delta_{v}=n(1-y/3), \quad \Delta[v']=d-\Delta_{v},
\nonumber \\
\Delta[\theta^{n}]=n\Delta_{\theta}=n(-1+y/6), \quad
\Delta[\theta']= d-\Delta_{\theta},
\nonumber \\
\Delta_{\omega} = 2-y/3,  \quad \Delta_{m}=1.
\label{Deltas}
\end{eqnarray}
These results are exact due to the exact expressions
$\gamma_{\nu}(g_{*},u_{*})=\gamma_{\kappa}(g_{*},u_{*})=y/3$ which follow
from the relations (\ref{beta}) at the fixed point.

\section{Inertial-range behaviour of the structure functions and the OPE}
\label{sec:OPE}

Solution of the RG equations gives the asymptotic expressions for the
various Green functions in the IR range, that is, for $\Lambda r \gg1$
and any fixed value of $mr$, where $\Lambda$ is the UV momentum scale from
(\ref{Lambda}) and $m$ is the IR scale from the correlators (\ref{2})
and (\ref{1.2}). In particular, for the structure functions (\ref{Ano})
one obtains:
\begin{equation}
S_{2n}(r) = D_{0}^{-n}\, r^{-2n\Delta_{\theta}}\, \xi_n (mr),
\label{strucS}
\end{equation}
cf. \cite{NSpass} for the scalar case. The inertial range corresponds to the
additional condition $mr\ll 1$. The asymptotic form of the scaling function
$\xi(mr)$ for small $mr$ is determined by OPE and has the form
\begin{equation}
\xi_{n}(mr)=\sum_{F} C_{nF}\,(mr)^{\Delta_{F}},
\label{OR}
\end{equation}
where $\Delta_{F}$ are the critical dimensions of the relevant composite
fields (composite operators) $F$ and $C_{nF}(mr)$ are coefficients regular
in $(mr)^{2}$.

Obviously, the leading terms of the asymptotic behaviour of the function
(\ref{OR}) for $mr\ll1$ are determined by the operators with smallest
dimensions $\Delta_{F}$. However, some additional considerations should
be taken into account. The operators whose dimensions appear in (\ref{OR})
are those allowed by the symmetry of the model and by the symmetry of the
quantity on left-hand side. In our model, these are scalar operators,
invariant with respect to Galilean transformation and with respect to the
shift $\theta\to\theta+{\rm const}$. The number of the fields $\theta$
in the operator $F$
cannot exceed their number on the left-hand side: this is a consequence
of the linearity of the original stochastic equation in $\theta$. The
operators which can be represented as total derivatives, $\partial F$,
have vanishing mean values and do not contribute to (\ref{OR}). For more
detailed discussion of all these points see e.g. \cite{NSpass}.

In models of critical behaviour (like e.g. the $\phi^{4}$ model)
the leading contribution to the OPE is given by the simplest operator $F=1$
with $\Delta_{F}=0$. The feature specific of the models of turbulence is
the existence of the so-called ``dangerous'' operators with {\it negative}
critical dimensions $\Delta_{F}<0$. They dominate the small-$mr$ asymptotic
behaviour of the scaling functions and lead to singular dependence of the
IR scale, that is, to the anomalous scaling \cite{RG,JphysA}.

Like in the scalar \cite{NSpass} and rapid-change vector \cite{vector} cases,
the most dangerous in our model are the simple operators $\theta$ whose
dimensions are known exactly, see (\ref{Deltas}). But they are not
invariant with respect to the shift and do not contribute to (\ref{OR}).
Thus the leading terms for $mr\ll1$  in (\ref{OR}) are given by the operators
with minimal possible number of spatial derivatives, which guarantee the
invariance with respect to that shift, that is, one derivative per
each field. They have the forms $\partial \theta \dots \partial \theta$
with even number of factors $\partial \theta$ (otherwise it is impossible
to get a scalar of the needed form; see below) and all possible contractions
of the vector indices of the fields and derivatives. For a scalar field
$\theta$ there is only one variant of the contraction:
$(\partial_{i}\theta\partial_{i}\theta)^{n}$, and $\Delta_{n}$ in
(\ref{HZ3}) is the critical dimension of this operator \cite{RG,NSpass}.

For the vector field $\theta$ the number of possible contraction variants
in the structure $(\partial\theta)^{2n}$ rapidly increases with $n$.
All the operators with a given $n$ mix heavily in renormalization, so that
the leading exponents $\Delta_{F}$ in (\ref{OR}) are not determined by an
individual operator, but rather are the eigenvalues of the matrix of critical
dimensions. The minimal eigenvalue gives the leading term of the small-$mr$
behaviour, and the others give the corrections which, for small $y$, can be
very close to the leading term.

For $n=1$ the contraction variant is still unique:
$\partial_{i}\theta _{j} \partial_{i}\theta_{j}$.
Like in the scalar case \cite{RG}, its dimension $\Delta_{n}=0$ is found
exactly from a certain Schwinger equation, so that the function
$S_{2}\propto r^{-2\Delta_{\theta}}$ reveals no anomalous scaling;
cf.~\cite{vector} for the rapid-change case. (The second variant
$\partial_{i}\theta _{j} \partial_{j} \theta_{i} =
\partial _{i} \partial_{j} (\theta _{j}\theta_{i})$ leads  to a total
derivative and thus gives no contribution to (\ref{OR})).
However, for $n=2$ there are already 7 variants \cite{LA,vector}, and for
$n=9$ as many as 47 246 \cite{Novikov}.

Furthermore, there are some nontrivial linear relations between the
operators with the same $n$, which reduce the number of independent
monomials, especially for low values of $d$ (up to 6 for $n=2$ and general
$d$ and up to 154 for $n=9$ and $d=3$), but give rise to a difficult
problem of finding and exclusion of all the redundant operators; see
\cite{Novikov} for a detailed analysis. In two dimensions, the transverse
vector field is expressed in terms of a scalar field by means of the
antisymmetric Levy--Civita tensor:
$\theta_{i} = \epsilon_{ik} \partial_{k} \psi$. This allows one to
diagonalize the matrix of critical dimensions of the operators
$(\partial\theta)^{2n}$ for arbitrary $n$ and to derive the explicit
results for the leading anomalous exponents to the order $\xi^2$
\cite{Novikov}.

For a general $d$, the families with different $n$ should be studied
separately. Surprisingly enough, the problem drastically simplifies for
$d\to\infty$, which allows one to achieve very high values of $n$ and
to obtain simple explicit results for leading and correction exponents
\cite{vector,matrix}. So far, all these results were confined with the
Kraichnan's velocity ensemble.

\section{Critical dimensions of the operators $(\partial\theta)^{2n}$
for large $d$.} \label{sec:d}

The analysis of the renormalization of the composite operators
$(\partial\theta)^{2n}$ and the practical first-order (one-loop)
calculation of the renormalization constants and critical dimensions in the
model (\ref{action}), (\ref{actionV}) is very similar to the case of vector
Kraichnan's model, which is discussed in great details in ref.~\cite{vector};
see especially sec.~B and C and appendix~B. The only relevant one-loop
Feynman diagrams in the two models differ only in the scalar factor stemming
from the integrals over the frequency; all the tensor factors (projectors,
vertices etc.) are exactly the same. As a consequence, the expressions for
the renormalization constants here can be obtained by the substitution
$g \bar S_{d} /\varepsilon \to g \bar S_{d} / yu(u+1)$ in expressions like
(6.19), (B15) and (B17) in \cite{vector}, while the final fixed-point
expressions for the critical dimensions are obtained by the replacement
$\varepsilon\to y/3$ in expressions like (6.24), (6.25) or (B6c)
($\varepsilon=\xi$ in the notation of \cite{vector}).

For this reason, below we only briefly discuss the renormalization of the
operators $(\partial\theta)^{2n}$ and the simplifications that occur for
large $d$. Detailed justification given in ref.~\cite{vector} for the
rapid-change vector model literally applies to the present case.

The critical dimension $\Delta$ of an arbitrary scalar composite operator
of the form $(\partial\theta)^{2n}$ in the first order of the expansion in
$y$ has the form $\Delta = \Delta_{1}(d) y + O(y^{2})$ with a certain
coefficient $\Delta_{1}(d)$ that depends on $d$. The first terms of its
expansion in $1/d$ have the forms
\[ \Delta_{1}(d) = 2k + \Delta_{11} /d  + O(1/d^{2}), \]
where $k$ is an integer number satisfying the inequalities
$0\le k\le n$ and $\Delta_{11}$ is a numerical coefficient independent of
$y$ and $d$. It turns out that in the first order in $1/d$ the subsets
with different $k$ ``decouples'' from one another, so that their
renormalization can be studied separately.

It is clear that for large $d$, dangerous operators with $\Delta_{11}<0$
can be present only in the subsets with $k=0$. They are formed by the
operators $(\partial\theta)^{2n}$ of a very special type, namely, those
in which all the fields are contracted only with the fields, and the
derivatives are contracted only with derivatives. For a given
$n$ all such operators are represented as the products
\begin{equation}
F = (\phi_{1})^{n_{1}}(\phi_{2})^{n_{2}} \dots (\phi_{q})^{n_{q}},
\label{Form}
\end{equation}
where $\sum_{k=1}^{q} kn_{k} = n$ and $\phi_{k}$ is a scalar operator
that includes $2k$ factors $\partial\theta$ and cannot be reduced to
a product of certain scalar factors. Such a basic factor can be written
in the form
\begin{equation}
\phi_{k} = \partial^{l_{1}} \theta_{s_{k}} \partial^{l_{1}}
\theta_{s_{1}}
\partial^{l_{2}}   \theta_{s_{1}}
\partial^{l_{2}}   \theta_{s_{2}}
\partial^{l_{3}}   \theta_{s_{2}}    \partial^{l_{3}} \theta_{s_{3}}
\cdots \partial^{l_{k}}\theta_{s_{k-1}} \partial^{l_{k}}
\theta_{s_{k}}.
\label{Form2}
\end{equation}

Let us give a few examples. For $n=2$ there are two operators of the
type (\ref{Form}):
\[ F= \{ \phi_{1}^{2}, \ \phi_{2} \}, \]
for $n=3$ there are three operators:
\[ F= \{ \phi_{1}^{3}, \ \phi_{1}\phi_{2}, \ \phi_{3}  \}, \]
for $n=4$ there are five operators:
\[ F= \{ \phi_{1}^{4}, \ \phi_{1}^{2}\phi_{2}, \
\phi_{2}^{2}, \  \phi_{1}\phi_{3}, \ \phi_{4}  \}, \]
and for $n$ between 5 and 11 the number of relevant operators equals to
7, 11, 15, 22, 30, 42 and 56, respectively. These number are much smaller
than the total numbers of the operators $(\partial\theta)^{2n}$ with a
given $n$: for example, 2 rather than 6 for $n=2$ and 30 rather than 47 246
for $n=9$. However, for $n=28$ there are as many as 3718 operators
of the type (\ref{Form}),  so that the problem remains highly nontrivial
even for $d \to \infty$. Let us give the whole set of the operators of
the type (\ref{Form}) for $n=7$:
\[ F= \{ \phi_{1}^{7}, \ \phi_{1}^{5}\phi_{2}, \ \phi_{1}^{3}\phi_{2}^{2}, \
\phi_{1}\phi_{2}^{3}, \ \phi_{1}^{4}\phi_{3}, \
\phi_{1}\phi_{3}^{2}, \ \phi_{1}^{3}\phi_{4}, \
\phi_{1}^{2}\phi_{5}, \]
\begin{equation}
\phi_{1}\phi_{6}, \ \phi_{1}^{2}\phi_{2}\phi_{3}, \
\phi_{2}^{2}\phi_{3}, \ \phi_{1}\phi_{2}\phi_{4}, \
\phi_{2}\phi_{5}, \  \phi_{3}\phi_{4}, \ \phi_{7} \}
\label{n=7}
\end{equation}
and for $n=8$:
\[ F= \{ \phi_{1}^{8}, \ \phi_{1}^{6}\phi_{2}, \
\phi_{1}^{4}\phi_{2}^{2}, \ \phi_{1}^{2}\phi_{2}^{3}, \
\phi_{2}^{4}, \ \phi_{1}^{5}\phi_{3}, \  \phi_{1}^{2}\phi_{3}^{2}, \]
\[ \phi_{2}\phi_{3}^{2}, \
\phi_{1}^{3}\phi_{2}\phi_{3}, \ \phi_{1}\phi_{2}^{2}\phi_{3}, \
\phi_{4}^{2}, \ \phi_{1}^{4}\phi_{4},
\phi_{1}^{2}\phi_{2}\phi_{4}, \
\phi_{2}^{2}\phi_{4}, \]
\begin{equation}
\phi_{1}\phi_{3}\phi_{4}, \ \phi_{1}^{3}\phi_{5}, \
\phi_{1}\phi_{2}\phi_{5}, \  \phi_{3}\phi_{5}, \
\phi_{1}^{2}\phi_{6}, \ \phi_{2}\phi_{6}, \
\phi_{1}\phi_{7}, \  \phi_{8} \}.
\label{n=8}
\end{equation}

Renormalization of the families of operators of the type (\ref{Form}) with
different $n$ can be studied separately: due to the linearity of the
original problem (\ref{1}), the operators $(\partial\theta)^{2n}$
do not admix in renormalization to the operators $(\partial\theta)^{2k}$
if $n>k$. The leading term of the double expansion in $y$ and $1/d$ for
the matrix of critical dimensions of the family (\ref{Form}) with a certain
given $n$ has the form
\begin{equation}
\Delta = - \frac{y}{3d} \, \widetilde\Delta + \dots,
\label{DeH}
\end{equation}
where the ellipsis stands for the corrections in $y$ and $1/d$ and
$\widetilde\Delta$ is a matrix with non-negative integer elements, which
are determined by certain simple rules \cite{vector}.

The diagonal element $\widetilde\Delta_{\alpha\alpha}$, corresponding to a
certain operator $F_{\alpha}$ with a given $n$, is given by the expression
\begin{equation}
\widetilde\Delta_{\alpha\alpha} = n-n_{1} -\sum_{k=2}^{q} n_{k} k(k-1),
\label{alik}
\end{equation}
where $n_{k}$ is the number of factors $\phi_{k}$ that constitute the
operator $F_{\alpha}$.

The non-diagonal elements are determined by the ``fusion and decay processes''
of the simple factors $\phi_{k}$. Let us choose in the operator $F_{\alpha}$
a pair of the simple factors $\phi_{k}$ and $\phi_{p}$ with certain $k$ and
$p$ ($k=p$ is allowed) and replace them by a single factor $\phi_{k+p}$.
Then the original operator $F_{\alpha}$ turns to some other $F_{\beta}$ with
the same $n$. This ``fusion process'' $\phi_{k}\phi_{p}\to \phi_{k+p}$ gives
to the matrix element $\widetilde\Delta_{\alpha\beta}$ a contribution of the
form $4kp$ with the summation over all possible pairs of factors
$\phi_{k}\phi_{p}$ entering into $F_{\alpha}$. For example, starting with
the operator $F_{1}=\phi_{1}^{3}$ one can get the operator
$F_{2}=\phi_{1}\phi_{2}$ by means of the fusion
$\phi_{1}\phi_{1}\to \phi_{2}$ with $k=p=1$; this gives the matrix element
$\widetilde\Delta_{12} =  4kp \, C_{3}^{2}  =12$, where the factor
$C_{3}^{2}=3$ arises as the number of possibilities to choose the pair
$\phi_{1}\phi_{1}$ from the three factors of the type $\phi_{1}$ in $F_{1}$.

Furthermore, let us choose in the operator $F_{\alpha}$ some simple factor
$\phi_{k}$ and replace it with the factor $\phi_{k-p}\phi_{p}$ with a certain
$1\le p \le (k-1)$. Then the new operator $F_{\beta}$ with the same $n$
results. This ``decay process'' $\phi_{k} \to \phi_{k-p}\phi_{p}$ gives
to the matrix element $\widetilde\Delta_{\alpha\beta}$ a contribution of the
form $2k$ per each factor $\phi_{k}$ entering into $F_{\alpha}$, if the
factors $\phi_{k-p}$ and $\phi_{p}$ are different, that is, $p \ne k-p$,
or $k$ per each factor $\phi_{k}$ if they are identical, that is, $p=k-p$.

Example: two possible decays in the operator $F_{1}=\phi_{4}$ ($k=4$) give
rise to the operators $F_{2}=\phi_{2}^2$ ($k=p-k=2$) and
$F_{3}=\phi_{1}\phi_{3}$ ($p=1$, $p-k=3$); the corresponding matrix elements
equal to $\widetilde\Delta_{12} = k = 4$ and $\widetilde\Delta_{13}=2k=8$.
Another example: the decay $\phi_{2} \to \phi_{1}\phi_{1}$ in the operator
$F_{1}=\phi_{2}^{3}$ gives rise to $F_{2}=\phi_{1}^2\phi_{2}^2$ with the
matrix element $\widetilde\Delta_{12} = 3k = 6$; the factor 3 accounts for
the presence of three monomials $\phi_{2}$ in the initial operator $F_{1}$.

If the operator $F_{\alpha}$ gives rise to another operator $F_{\beta}$
as a result of certain fusion $\phi_{k}\phi_{p}\to \phi_{k+p}$, then it
is clear that $F_{\beta}$ gives rise to  $F_{\alpha}$ as a result of the
``inverse decay'' $\phi_{k+p}\to\phi_{k}\phi_{p}$. Thus the matrix elements
$\widetilde\Delta_{\alpha\beta}$ and $\widetilde\Delta_{\beta\alpha}$ can
vanish only simultaneously (which happens very often); otherwise they both
are not equal to zero (but, in general, are not equal to each other).

\newpage

For $n=2$ and 3 the matrices $\widetilde\Delta$ have the forms:
\begin{equation}
\left( \matrix { 0 & 4 \cr  2 & 0 \cr } \right), \quad
\left( \matrix { 0 & 12 & 0 \cr 2 & 0 & 8 \cr 0 & 6 & 3 \cr  } \right).
\label{Mat23}
\end{equation}

For the families of operators that mix in renormalization the exponents
$\Delta_{F}$ in (\ref{OR}) are determined by the eigenvalues of the matrices
$\widetilde\Delta$ with the subsequent substitution into (\ref{DeH}). The
eigenvalues of the matrices (\ref{Mat23}) for $n=2$ are $\pm 2\sqrt{2}$,
while for $n=3$ they are equal to $1+ 10\cos\psi = 9,673557$,
$1- 5 \cos\psi -5\sqrt{3} \sin\psi = - 7,64689$,
$1- 5 \cos\psi + 5\sqrt{3} \sin\psi = 0,973333$, where we have denoted
$\psi=  (1/3) {\rm arctg} \left( 6 \sqrt{434} \right)$.
For higher $n$ the eigenvalues were found numerically.

\bigskip

The matrices $\widetilde\Delta$ for $n$ as high as 6 can be found in
\cite{vector}. As two more examples, here we give the matrices for
the case $n=7$:
$$ \left( \matrix{
    0 & 84 & 0 & 0 & 0 & 0 & 0 & 0 & 0 & 0 & 0 & 0 & 0 & 0 & 0 \cr
    2 &  0 & 40 & 0 & 40 & 0 & 0 & 0 & 0 & 0 & 0 & 0 & 0 & 0 & 0 \cr
    0 &  4 & 0 & 12 & 0 & 0 & 16 & 0 & 0 & 48 & 0 & 0 & 0 & 0 & 0 \cr
    0 & 0 & 6 & 0 & 0 & 0 & 0 & 0 & 0 & 0 & 24 & 48 & 0 & 0 & 0 \cr
    0 & 6 & 0 & 0 & 3 & 0 & 48 & 0 & 0 & 24 & 0 & 0 & 0 & 0 & 0 \cr
    0 & 0 & 0 & 0 & 0 & 6 & 0 & 0 & 36 & 12 & 0 & 0 & 0 & 24 & 0  \cr
    0 & 0 & 4 & 0 & 8 & 0 & 8 & 48 & 0 & 0 & 0 & 12 & 0 & 0 & 0 \cr
    0 & 0 & 0 & 0 & 0 & 0 & 10 & 15 & 40 & 10 & 0 & 0 & 4 & 0 & 0 \cr
    0 & 0 & 0 & 0 & 0 & 6 & 0 & 12 & 24 & 0 & 0 & 12 & 0 & 0 & 24 \cr
    0 & 0 & 6 & 0 & 2 & 16 & 0 & 24 & 0 & 3 & 4 & 24 & 0 & 0 & 0 \cr
    0 & 0 & 0 & 6 & 0 & 0 & 0 & 0 & 0 & 4 & 3 & 0 & 48 & 16 & 0  \cr
    0 & 0 & 0 & 4 & 0 & 0 & 2 & 0 & 32 & 8 & 0 & 8 & 16 & 8 & 0  \cr
    0 & 0 & 0 & 0 & 0 & 0 & 0 & 2 & 0 & 0 & 10 & 10 & 15 & 0 & 40  \cr
    0 & 0 & 0 & 0 & 0 & 8 & 0 & 0 & 0 & 0 & 4 & 6 & 0 & 11 & 48  \cr
    0 & 0 & 0 & 0 & 0 & 0 & 0 & 0 & 14 & 0 & 0 & 0 & 14 & 14 & 35 \cr
             } \right) $$

\noindent and for $n=8$:

\newpage

{\small  $$
\left( \matrix{
0 & 112 & 0 & 0 & 0 & 0 & 0 & 0 & 0 & 0 & 0 & 0 & 0 & 0 & 0 & 0 & 0 & 0 & 0 & 0 & 0 & 0 \cr
2 & 0 & 60 & 0 & 0 & 48 & 0 & 0 & 0 & 0 & 0 & 0 & 0 & 0 & 0 & 0 & 0 & 0 & 0 & 0 & 0 & 0 \cr
0 & 4 & 0 & 24 & 0 & 0 & 0 & 0 & 64 & 0 & 0 & 16 & 0 & 0 & 0 & 0 & 0 & 0 & 0 & 0 & 0 & 0 \cr
0 & 0 & 6 & 0 & 4 & 0 & 0 & 0 & 0 & 48 & 0 & 0 & 48 & 0 & 0 & 0 & 0 & 0 & 0 & 0 & 0 & 0 \cr
0 & 0 & 0 & 8 & 0 & 0 & 0 & 0 & 0 & 0 & 0 & 0 & 0 & 96 & 0 & 0 & 0 & 0 & 0 & 0 & 0 & 0 \cr
0 & 6 & 0 & 0 & 0 & 3 & 0 & 0 & 40 & 0 & 0 & 60 & 0 & 0 & 0 & 0 & 0 & 0 & 0 & 0 & 0 & 0 \cr
0 & 0 & 0 & 0 & 0 & 0 & 6 & 4 & 12 & 0 & 0 & 0 & 0 & 0 & 48 & 0 & 0 & 0 & 36 & 0 & 0 & 0 \cr
0 & 0 & 0 & 0 & 0 & 0 & 2 & 6 & 0 & 12 & 0 & 0 & 0 & 0 & 0 & 0 & 0 & 48 & 0 & 36 & 0 & 0 \cr
0 & 0 & 6 & 0 & 0 & 2 & 24 & 0 & 3 & 12 & 0 & 0 & 36 & 0 & 0 & 24 & 0 & 0 & 0 & 0 & 0 & 0 \cr
0 & 0 & 0 & 6 & 0 & 0 & 0 & 16 & 4 & 3 & 0 & 0 & 0 & 12 & 16 & 0 & 48 & 0 & 0 & 0 & 0 & 0 \cr
0 & 0 & 0 & 0 & 0 & 0 & 0 & 0 & 0 & 0 & 16 & 0 & 0 & 8 & 16 & 0 & 0 & 0 & 0 & 0 & 0 & 64 \cr
0 & 0 & 4 & 0 & 0 & 8 & 0 & 0 & 0 & 0 & 0 & 8 & 24 & 0 & 0 & 64 & 0 & 0 & 0 & 0 & 0 & 0 \cr
0 & 0 & 0 & 4 & 0 & 0 & 0 & 0 & 8 & 0 & 0 & 2 & 8 & 4 & 16 & 0 & 32 & 0 & 32 & 0 & 0 & 0 \cr
0 & 0 & 0 & 0 & 4 & 0 & 0 & 0 & 0 & 8 & 16 & 0 & 4 & 8 & 0 & 0 & 0 & 0 & 0 & 64 & 0 & 0 \cr
0 & 0 & 0 & 0 & 0 & 0 & 8 & 0 & 0 & 4 & 12 & 0 & 6 & 0 & 11 & 0 & 0 & 16 & 0 & 0 & 48 & 0 \cr
0 & 0 & 0 & 0 & 0 & 0 & 0 & 0 & 10 & 0 & 0 & 10 & 0 & 0 & 0 & 15 & 12 & 0 & 60 & 0 & 0 & 0 \cr
0 & 0 & 0 & 0 & 0 & 0 & 0 & 0 & 0 & 10 & 0 & 0 & 10 & 0 & 0 & 2 & 15 & 8 & 0 & 20 & 40 & 0 \cr
0 & 0 & 0 & 0 & 0 & 0 & 0 & 10 & 0 & 0 & 0 & 0 & 0 & 0 & 10 & 0 & 6 & 18 & 0 & 0 & 0 & 60 \cr
0 & 0 & 0 & 0 & 0 & 0 & 6 & 0 & 0 & 0 & 0 & 0 & 12 & 0 & 0 & 12 & 0 & 0 & 24 & 4 & 48 & 0 \cr
0 & 0 & 0 & 0 & 0 & 0 & 0 & 6 & 0 & 0 & 0 & 0 & 0 & 12 & 0 & 0 & 12 & 0 & 2 & 24 & 0 & 48 \cr
0 & 0 & 0 & 0 & 0 & 0 & 0 & 0 & 0 & 0 & 0 & 0 & 0 & 0 & 14 & 0 & 14 & 0 & 14 & 0 & 35 & 28 \cr
0 & 0 & 0 & 0 & 0 & 0 & 0 & 0 & 0 & 0 & 8 & 0 & 0 & 0 & 0 & 0 & 0 & 16 & 0 & 16 & 16 & 48 \cr
  } \right) $$}

In the both cases the operators are numbered according to their order in
(\ref{n=7}) and (\ref{n=8}).

The algorithm described above for constructing the matrices
$\widetilde\Delta$ was realized as a computer program; using it we have
found all the matrices and their eigenvalues up to the family with $n=28$
(it involves 3718 relevant operators). The computations for larger $n$
become too time consuming (mostly for finding the eigenvalues) but they
appear unnecessary: the eigenvalues demonstrate interesting regularities,
which allows one to suggest for them some simple ``empiric'' formulae
that become nearly exact for $n$ large enough.

\section{Critical dimensions of the operators $(\partial\theta)^{2n}$
for large $n$.} \label{sec:n}

The analysis of the matrices $\widetilde\Delta$ with $n\le28$ reveals the
following properties.

All the matrices can be brought to diagonal form; all their eigenvalues are
real and differ from zero. (One can check that the low-order matrices
$\widetilde\Delta$ can be made symmetric by a proper normalization of the
basis operators (\ref{Form}); this is likely true for all orders. However,
the elements of the matrices become non-integer after that procedure.)
Among the eigenvalues for a given $n$ there are always positive and negative
ones. Starting from $n=3$, the number of positive eigenvalues exceeds the
number of negative eigenvalues (roughly speaking, twice). The maximum (by
modulus) positive and negative eigenvalues grow monotonously with $n$, and
the maximal positive eigenvalue is always larger (by modulus) than the
negative one (roughly speaking, twice for large $n$).
Let us give a few examples.

For $n=28$ there are 3718 operators of the form (\ref{Form}), 2569 positive
eigenvalues and 1149 negative ones. The maximal positive eigenvalue of the
matrix $\widetilde\Delta$ equals to 1484.5, the minimal one is $-782.1$.
For smaller values of $n$  the same five numbers are:
\begin{eqnarray}
1575;\quad 1072;\quad 503;\quad 1080.5;\quad -574.12 \quad  {\rm for}\ n=24,
\nonumber \\
1958;\quad 1337;\quad 621;\quad 1175.5;\quad -623.11 \quad  {\rm for}\ n=25,
\nonumber \\
2436;\quad 1674;\quad 762;\quad 1274.5;\quad -674.11 \quad  {\rm for}\ n=26,
\nonumber \\
3010;\quad 2070;\quad 940;\quad 1377.5;\quad -727.11 \quad  {\rm for}\ n=27.
\label{five}
\end{eqnarray}

The relation (\ref{DeH}) shows that dangerous operators with negative
critical dimensions correspond to positive eigenvalues of the matrices
$\widetilde\Delta$, which therefore are the most interesting ones.
Furthermore, the maximal (for a given $n$) eigenvalue determines the
leading term in the IR asymptotic behaviour of the structure function
$S_{2n}$ in (\ref{strucS}), (\ref{OR}), that is, the principal anomalous
exponent. The other eigenvalues determine corrections to the leading term;
they diverge for $mr\to0$ if the corresponding eigenvalue is positive and
decrease if it is negative.

The operators that possess definite critical dimensions (``scaling
operators'') are certain linear combinations of the basis monomials
(\ref{Form}). It turns out that the scaling operators corresponding to the
maximal eigenvalues involve all the monomials from the family with the given
$n$; all the coefficients are positive and become closer to each other as
$n$ grows. (All the other scaling operators necessarily involve
coefficients with different signs, because the eigenvectors of the matrix
$\widetilde\Delta$ in its symmetric form must be orthogonal.)
This ``democracy of monomials'' can be opposed to the two-dimensional
case, where the principal eigenvalues correspond to the operators of a
very special form: powers of the local dissipation rate
$(\partial_{i}\theta_{k}\partial_{i}\theta_{k}
-\partial_{i}\theta_{k}\partial_{k}\theta_{i})$; see \cite{Novikov}.

Let us give all maximal positive eigenvalues $\lambda_{0} (n)$ of the
matrices $\widetilde\Delta$ for all $n$ from 2 to~28:
\[ 2.828;\quad 9.67356;\quad 20.617;\quad 35.5888;\quad 54.5717;\quad
77.5602; \]
\[ 104.5518;\quad  135.55;\quad 170.54059;\quad 209.5366;\quad
252.5334; \]
\[ 299.53063;\quad 350.52832;\quad  405.53;\quad
464.5246;\quad  527.52308; \]
\[ 594.52175;\quad 665.52055;\quad 740.51949;\quad
819.51852;\quad 902.51765; \]
\begin{equation}
989.51686;\quad 1080.5;\quad 1175.5;\quad
1274.5;\quad 1377.5;\quad 1484.5.
\label{lambda}
\end{equation}

The figure~1 shows that the eigenvalues $\lambda_{0} (n)$, plotted against
the number $n$, are approximated nicely by a smooth curve (the upper solid
line). Surprisingly enough, that curve is described very well by a simple
analytic expression:
\begin{equation}
\lambda_{0} (n) = 2n^{2} - 3n +1/2 +O(1/n),
\label{kurv}
\end{equation}
where the $O(1/n)$ correction appears rather small already for $n$ not too
large. Indeed, the inspection of the maximal eigenvalues $\lambda_{0}(n)$
given in (\ref{lambda}) shows that the expression $2n^{2} -3n$ gives
exactly (and with no exceptions) their integer parts, and the refined
expression $2n^{2} -3n+1/2$ gives, starting from $n=5$, the first number
after the decimal point (it is equal to 5 for all $n\ge5$).

The next-to-maximal eigenvalues also form a smooth curve, the next ones form
their own branch, etc. All the branches are described by simple explicit
formulae, which rapidly become nearly exact as $n$ grows. Let us give
them for a few branches closest to the leading one (\ref{kurv}):
\begin{eqnarray}
\lambda_{0} (n) &=& 2n^{2} - 3n +1/2,  \nonumber \\
\lambda_{1}   (n) &=& 2n^{2} -  7 n + 7/2,   \nonumber \\
\lambda_{2,1} (n) &=& 2n^{2} - 11 n + 40/3,  \nonumber \\
\lambda_{2,2} (n) &=& 2n^{2} - 11 n + 23/3,  \nonumber \\
\lambda_{3,1} (n) &=& 2n^{2} - 15 n + 187/6,  \nonumber \\
\lambda_{3,2} (n) &=& 2n^{2} - 15 n + 135/6,  \nonumber \\
\lambda_{3,3} (n) &=& 2n^{2} - 15 n + 83/6,  \nonumber \\
\lambda_{4,1} (n) &=& 2n^{2} - 19 n + 57.1,  \nonumber \\
\lambda_{4,2} (n) &=& 2n^{2} - 19 n + 44.3,
\label{kurv1}
\end{eqnarray}
with corrections of order $O(1/n)$.

\begin{figure}
\begin{center}
\includegraphics[width=15cm]{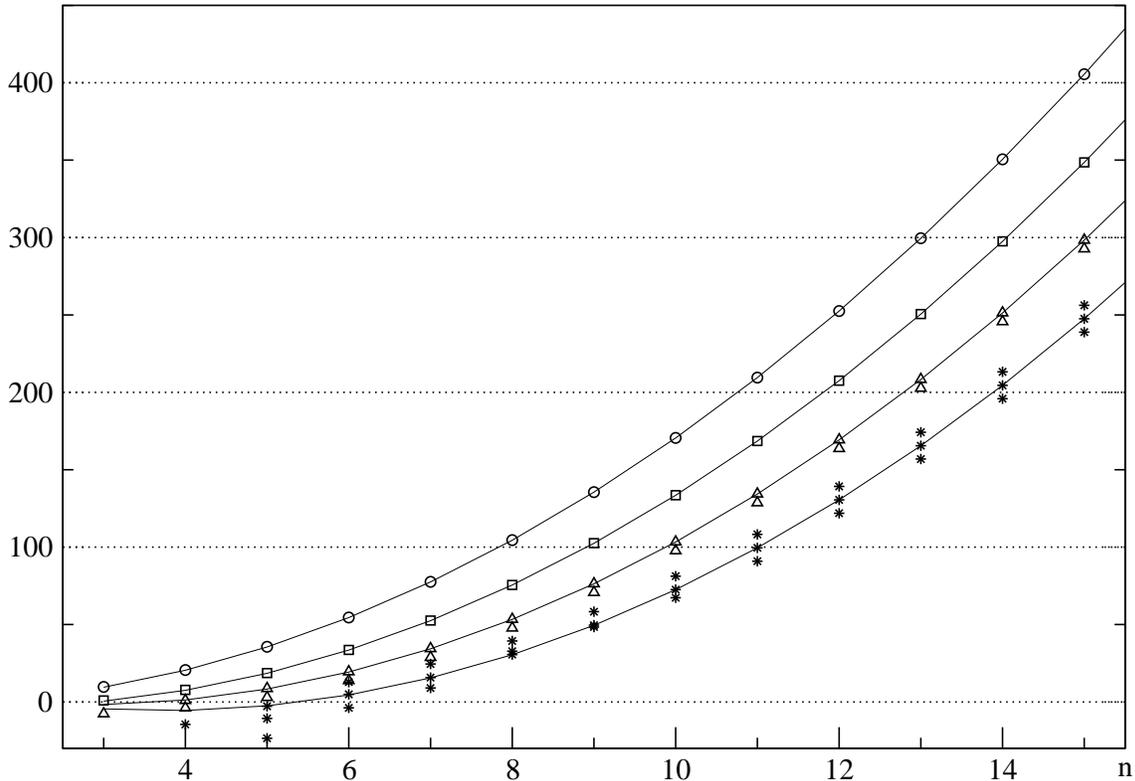}
\caption{\label{fig:Ev}
The eigenvalues of the matrices $\widetilde\Delta$. The circles correspond
to the leading branch $\lambda_{0}$ from (\ref{kurv}), the squares, triangles
and asterisks correspond to the branches $\lambda_{1}$, $\lambda_{2*}$ and
$\lambda_{3*}$ from (\ref{kurv1}). }
\end{center}
\end{figure}

It is interesting to note that, for $n=1$, the only eigenvalue $\lambda(1)=0$
(which is known exactly) belongs to neither curve. Probably this fact is
related to the observation that the agreement between the formulae
(\ref{kurv}), (\ref{kurv1}) and the exact numerical values for the
eigenvalues improves if the correction term is written in the form
$O(1/(n-1))$.

In their turn, the expressions (\ref{kurv}), (\ref{kurv1}) demonstrate
interesting regularities: the leading (quadratic in $n$) term is the same
for all the branches; the next-to leading term (linear in $n$) is negative
with the odd coefficient growing with the step of 4. The branches can be
grouped according the value of that coefficient, and the number of branches
with a given value grows: there is one branch in the first two groups, two
branches in the third group, three branches in the fourth group etc. It is
also worth noting that if the constant (independent of $n$) terms in
(\ref{kurv}) are replaced with the closest integer numbers and the $O(1/n)$
corrections are neglected, the resulting expressions give exactly the integer
parts for all the eigenvalues.

These regularities are also illustrated by figure~1, where all the positive
eigenvalues of the matrices $\widetilde\Delta$ are shown with $n$ from 3 to
15; according to (\ref{DeH}), they correspond to ``dangerous'' composite
operators with negative critical dimensions. The solid lines correspond to
representatives of the principal branches: $\lambda_{0}$ from
(\ref{kurv}) and $\lambda_{1}$, $\lambda_{2,1}$,  $\lambda_{3,3}$ from
(\ref{kurv1}). They are  plotted according the formulae (\ref{kurv}) and
(\ref{kurv1}) neglecting the $O(1/n)$ corrections. The other branches from
the groups $\lambda_{2,*}$ and $\lambda_{3,*}$ are not shown by solid lines
in order to make the picture more graspable. The circles denote the maximal
eigenvalues of the matrices $\widetilde\Delta$, and the squares denote the
next-to-maximal ones; they lie exactly on the principal branches
$\lambda_{0}$ and $\lambda_{1}$ from (\ref{kurv}), (\ref{kurv1}).
The eigenvalues that correspond to the two branches of the group
$\lambda_{2,*}$ are denoted by triangles and the eigenvalues of the
three branches of the next group $\lambda_{3,*}$ are denoted by asterisks.

The general formulae (\ref{kurv}), (\ref{kurv1}) for negative eigenvalues
do not work so well in comparison with positive ones. This fact is
illustrated by the same figure~1, where some negative eigenvalues
(with $n$ from 4 to 6) are also shown.
It turns out, however, that the negative eigenvalues form their own
pronounced branches; the principal one is described by the empiric formula
\begin{eqnarray}
\lambda_{0}^{(-)} (n) = - n^{2} + 2 + O(1/n).
\label{kurv3}
\end{eqnarray}
Comparison of expressions (\ref{kurv}) and (\ref{kurv3}) shows that the
ratio of the maximal positive and maximal (by modulus) negative eigenvalues
of the matrix $\widetilde\Delta$ tends to 2 as $n$ grows, in agreement with
the numerical values for $24 \le n \le 28$ given in (\ref{lambda}).

\section{Conclusion} \label{sec:Conc}

By means of the field theoretic renormalization group and operator product
expansion we studied the problem of anomalous scaling of the transverse
vector field passively advected by a turbulent velocity field. The dynamics
of the vector field is governed by the stochastic equation (\ref{1}),
(\ref{2}), while the velocity was described by the stirred NS equation
(\ref{1.1}), (\ref{1.2}), (\ref{1.9}). The anomalous scaling arises as a
consequence of the existence in the OPE of the so-called ``dangerous''
composite fields (operators) with negative critical dimensions.
The leading terms of the inertial-range asymptotic behaviour of the
structure functions (\ref{strucS}), (\ref{OR}) are determined by the
matrices of critical dimensions of the families of composite fields
of the form $(\partial\theta\partial\theta)^{2n}$. For $d\to\infty$,
dangerous operators can be present only in the subsets of operators of
the special form, (\ref{Form}), (\ref{Form2}), and the corresponding
matrices of critical dimensions can be constructed by means of a simple
algorithm. This allowed us to calculate them in the leading order of the
double expansion in $y$ and $1/d$ up to the order $n=28$. The eigenvalues
of those matrices (that is, the critical dimensions of the corresponding
families of operators) demonstrate intriguing pronounced regularities.
This fact allows one to describe them by simple empiric formulae which
become practically exact as $n$ grows. In particular, the leading term of
the asymptotic behaviour of the structure function (\ref{strucS}) in the
inertial range has the form
\begin{eqnarray}
S_{2n}({\bf r}) \simeq D_{0}^{-n}\,  r^{n(2-y/3)}\, (mr)^{\Delta_{n}},
\nonumber \\
\Delta_{n} = - (y/3d) \left( 2n^{2} - 3n +1/2 + O(1/n)\right),
\label{thule}
\end{eqnarray}
there are also explicit expressions for the correction exponents. Thus,
the complete description of the anomalous scaling for our vector model
is given for all $n$.

The regularities demonstrated by the critical dimensions of the sets of
composite fields and their branches, discussed in the present paper, are so
intriguing that we cannot but think that some unknown symmetry lies behind
them. One may think that understanding the relation between the anomalous
scaling, statistical conservation laws and operator product expansion will
be useful here; see \cite{AnPa} for the scalar Kraichnan's case.
In this connection it is interesting to note that the critical
dimensions of certain composite operators in quantum chromodynamics (QCD)
and in the $N=4$ supersymmetric gauge theories also show interesting
behaviour (also in the large-$n$ limit and in the one-loop approximation):
in particular, for the QCD case the corresponding evolution equations appear
equivalent to the integrable Heisenberg model \cite{Derk}.

The Kraichnan's rapid-change model of passive scalar advection is sometimes
referred to as ``the Ising model of fluid turbulence.'' In this connection,
it is worth recalling that the original Ising (or, better to say,
Lentz--Ising) model of magnetism, first introduced in 1920 \cite{Li2}, has
still remained a source of inspiration for new physical and mathematical
ideas and techniques like integrability, fermion--boson transformations,
conformal invariance and discrete holomorphisity: for a recent discussion,
see \cite{CIDHI} and references therein.

One may think that, in spite of a great deal of work devoted to Kraichnan's
model and its descendants, the deep physical and mathematical contents that
lie behind them are not completely disclosed. We believe that identifying the
hypothetical symmetry of the passive vector problem that gives rise to the
intriguing regularities discussed in the present paper will help to reach a
deeper understanding of the anomalous scaling in the real fluid turbulence.

\section*{Acknowledgements}

The authors thank S.\'E. Derkachev, Michal Hnatich, Juha Honkonen and Paolo
Muratore Ginanneschi for discussions.
The work was supported in part by the Russian Foundation for Fundamental
Research (project~12-02-00874-a).

\end{document}